\documentclass{emulatemyapj}
\pdfoutput=1
\usepackage[colorlinks=true,breaklinks=true,citecolor=blue,linkcolor=blue,urlcolor=blue]{hyperref}

\begin{document}

\shorttitle{Discrete steps in dispersion measures of Fast Radio Bursts}
\shortauthors{Michael Hippke}
\title{Discrete steps in dispersion measures of Fast Radio Bursts}

\author{Michael Hippke}
\email{hippke@ifda.eu}
\affil{Institute for Data Analysis, Luiter Stra{\ss}e 21b, 47506 Neukirchen-Vluyn, Germany}

\author{Wilfried F. Domainko}
\email{wilfried.domainko@mpi-hd.mpg.de}
\affil{Max-Planck-Institut f¨ur Kernphysik, P.O. Box 103980, D-69029 Heidelberg, Germany}

\author{John G. Learned}
\email{jgl@phys.hawaii.edu}
\affil{High Energy Physics Group, Department of Physics and Astronomy, University of Hawaii, Manoa 327 Watanabe Hall, 2505 Correa Road Honolulu, Hawaii 96822 USA}

\begin{abstract}
Fast Radios Bursts (FRBs) show large dispersion measures (DMs), suggesting an extragalactic location. We analyze the DMs of the 11 known FRBs in detail and identify steps as integer multiples of half the lowest DM found, 187.5cm$^{-3}$ pc, so that DMs occur in groups centered at 375, 562, 750, 937, 1125cm$^{-3}$ pc, with errors observed $<$5\%. We estimate the likelhood of a coincidence as $5:10,000$. We speculate that this could originate from a Galaxy population of FRBs, with Milky Way DM contribution as model deviations, and an underlying generator process that produces FRBs with DMs in discrete steps. However, we find that FRBs tend to arrive at close to the full integer second, like man-made perytons. If this holds, FRBs would also be man-made. This can be verified, or refuted, with new FRBs to be detected.
\end{abstract}

\section{Introduction}
Fast Radio Bursts (FRBs) are bright ($\sim$Jy peak flux densities) millisecond-duration pulses that have dispersion measures (DMs) higher than those expected from sources within the Galaxy. Eleven FRBs have been detected so far, nine with the Parkes and one at the Arecibo telescope. The DMs, as summarized in Table~\ref{tab:dms}, are in between 375 and 1103cm$^{-3}$ pc, exceeding values expected from a Galaxy origin by a factor 3--20 \citep{Cordes2003}. The large values have been attributed to emission stellar coronae \citep{Loeb2014} or ionized nebulae \citep{Kulkarni2014}. However, as pointed out by \citet{Ravi2015}, these explanations are inconsistent as there is a ``lack of observed deviations from the cold plasma dispersion law'' \citep{Tuntsov2014,Katz2014,Dennison2014}; and from constraints on H$\alpha$ observations on dense ionized nebulae \citep{Kulkarni2014}.

If FRBs were ``standard candles'', they could be used as cosmological tools to constrain the equation of state of dark energy \citep{Zhou2014}. It is therefore important to learn about their location. We have asked the question: What can we learn from the DM values of the eleven known FRBs? If some would originate from within our Galaxy, while others would have traveled through the cold plasma of one (or more) distant other galaxies plus the intergalactic medium, then we might see a clustering in their DMs, perhaps in distinct groups. We will explain our statistical methods in section~\ref{sec:method}, and present the results in section~\ref{sec:results}.

\section{Method}
\label{sec:method}
The small number of only 11 FRBs poses a great danger: Fitting a linear model leaves only $n-1$ free variables. Furthermore, there is an indefinite number of fits and groups one can try, e.g. polynomial trends, power-law-fits and so on. Consequently, there is an indefinite number of \textit{perfect} fits (but most are expected to be exotic). This is in principle true in \textit{all} data analysis, if one does not have an \textit{a priori} hypothesis. We note that we have no hypothesis for the distribution, it is simply a possibly true observation. After identifying this concern, we decide to restrict ourselves to only the most simple data treatments, such as two groups, or a linear fit, in the present data of only eleven FRBs.

In order to test the presence of two distinct groups, we defined a \textit{lower} and a \textit{higher} group. Then, we attributed the eleven FRBs to either group, with a sliding divider. If two groups were to explain the distribution, we would expect a decrease in standard deviation from the group mean, in the correct attribution of the FRBs to two groups.

\begin{table*}
\small
\caption{All FRBs known, sorted by their DMs\label{tab:dms}, and model results}
\begin{tabular}{llllll}
\tableline
FRB  & Telescope & DM (cm$^{-3}$ pc) & DM est. &  DM dev. & Source \\ 
\tableline
010724 & Parkes  & 375$\pm1$       &  375     &                 &\citet{Lorimer2007} \\
120127 & Parkes  & 553.3$\pm0.3$   &  562.5   & +9.2 (+1.6\%)   &\citet{Thornton2013} \\
121102 & Arecibo & 557$\pm2$       &  562.5   & +5.1 (+0.9\%)   &\citet{Spitler2014} \\
140514 & Parkes  & 562.7$\pm0.6$   &  562.5   & -0.2 (-0.04\%)  &\citet{Petroff2015} \\
110627 & Parkes  & 723.3$\pm0.3$   &  750     & +27  (+3.6\%)   &\citet{Thornton2013} \\
010621 & Parkes  & 746$\pm1$       &  750     & +4   (+0.5\%)   &\citet{Keane2012} \\
131104 & Parkes  & 779$\pm0.3$     &  750     & -29  (-3.9\%)   &\citet{Ravi2015} \\
011025 & Parkes  & 790$\pm3$       &  750     & -40  (-5.3\%)   &\citet{Burke2014} \\
110220 & Parkes  & 944.38$\pm0.05$ &  937.5   & -6.9 (-0.7\%)   &\citet{Thornton2013} \\
110703 & Parkes  & 1103.7$\pm0.7$  &  1125    & +21.4 (+1.9\%)  &\citet{Thornton2013} \\
121002 & Parkes  & 1628.76$\pm0.05$&  1687.5  & +89 (+3.5\%)    &\citet{Thornton2013PhD} \\
\tableline
\multicolumn{5}{l}{Uncertainties are as summarized in \citet{Keane2015}}
\end{tabular}
\end{table*}

\begin{figure}
\includegraphics[width=\linewidth]{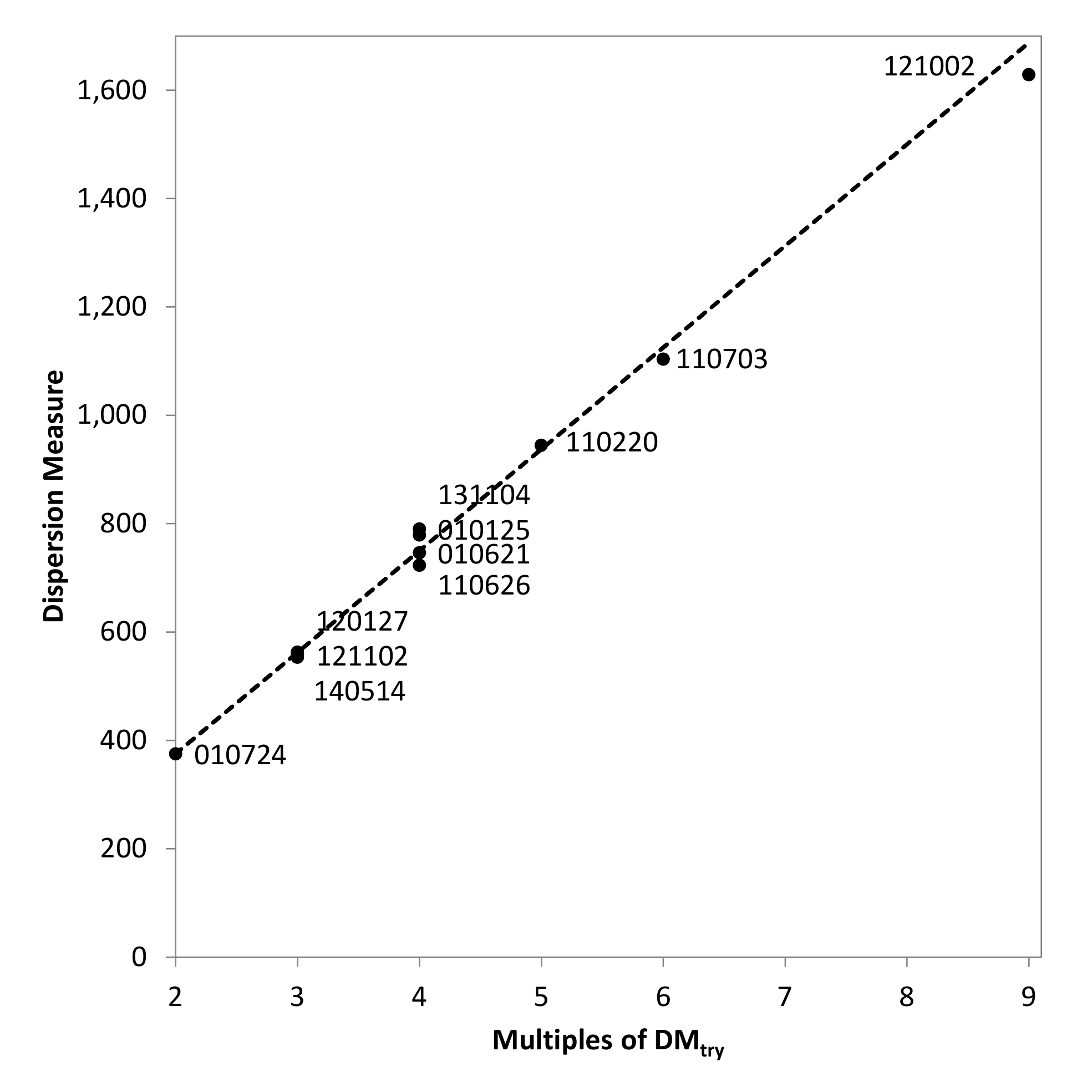}
\caption{\label{fig:dm}DM versus multiple of DM$_{try}$=187.5cm$^{-3}$ pc for the eleven FRBs.\\}
\end{figure}

\begin{figure}
\includegraphics[width=\linewidth]{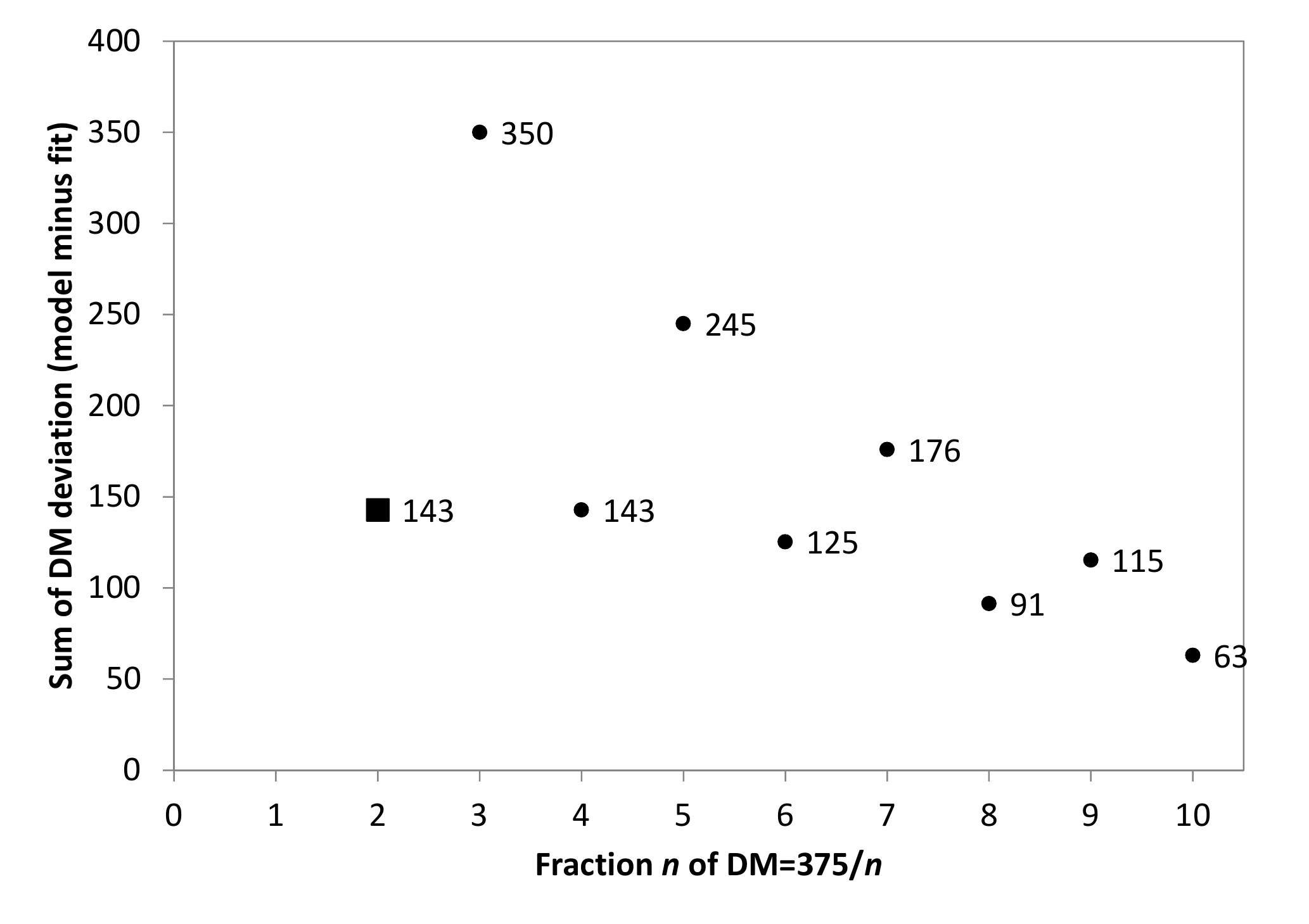}
\caption{\label{fig:dmcompare}Model comparison for linear fit of DM=$2/375$cm$^{-3}$ pc (square symbol), compared to other fractions.}
\end{figure}

\section{Results}
\label{sec:results}

\subsection{DM spacings}
After trying the sliding group attribution for all eleven FRBs, we have quickly concluded that in no case this produces a significant decrease in the groups' standard deviations to the group mean. We have, however, noticed a peculiarity. FRB010724, the first found, has the lowest DM of 375cm$^{-3}$ pc. Taking half this value, DM$_{try}=$187.5cm$^{-3}$ pc, gives near-perfect estimates for the other FRBs when multiplying DM$_{try}$ with integer numbers (see Table~\ref{tab:dms} and Fig.~\ref{fig:dm}). This linear trend is virtually unchanged when fitting a line through the highest and lowest DM data point: DM$_{try}$=184.3. 

It is important to note that we did not ``optimize'' this fit. To be precise, we can use three methods to create Fig.~\ref{fig:dm}. The first is to put a straight line through the two FRBs with the lowest, and the highest DM. This results in a slope of 184.3cm$^{-3}$ pc. The other FRBs are then scaled on the X-axis with their DM rounded to the nearest integer of the fit, i.e. FRB140514 has DM=562.7cm$^{-3}$ pc. This DM, divided by the slope of 184.3cm$^{-3}$ pc, gives 3.05. We round this to the nearest integer, 3. The deviations for all FRBs from this model fit are in Table~\ref{tab:dms}. 

The second method is to use the FRB with the smallest DM. This is FRB010724 which has DM=375cm$^{-3}$ pc. As explained above, we take half of this value as DM$_{try}$=187.5cm$^{-3}$ pc and multiply it with 2, 3, and the following integers. The other FRBs are then scaled on the X-axis with their DM rounded to the nearest integer of the fit, as described above.

Finally, one might use a least-squares fit to all data points, with FRBs attributed to groups as discussed. All methods give virtually identical results, as their slope parameters are very similar: 187.5cm$^{-3}$ pc (half the lowest FRB's DM), 184.3cm$^{-3}$ pc (highest/lowest) and 187.7cm$^{-3}$ pc (least squares fit). The coefficient of determination for these linear models is $R^2\sim0.99$.

A valid objection is that other slopes, such as $1/4$ of the base DM=375cm$^{-3}$ pc, would give even better fits. To quantify this, we have calculated the summed residuals of all such fractions 2..10 and show the result in Fig.~\ref{fig:dmcompare}. It can be clearly seen that the series $1/2, 1/4, 1/6...$ gives increasingly better fits. As $1/2$ is the simplest choice, and we are dealing with the problem of \textit{a posteriori} statistics, we will choose this value for further discussion.

As such fit quality can happen by coincidence, we have estimated its likelihood by generating 10,000 Monte Carlo runs for eleven DMs each in the range [300,2000]. For every sample of eleven randomly chosen DMs, we have checked how many gave a better fit for the linear trend with half the base value, or for a linear trend through the highest and lowest data points. The result is a striking 1:10,000 chance for the first option, and 5:10,000 for the second. In the random numbers, we also see average summed DM residuals $>$500, as expected for the series following $1/3, 1/5, 1/7...$ when approaching $1/2$ (Fig.~\ref{fig:dmcompare}).

\begin{figure}
\includegraphics[width=\linewidth]{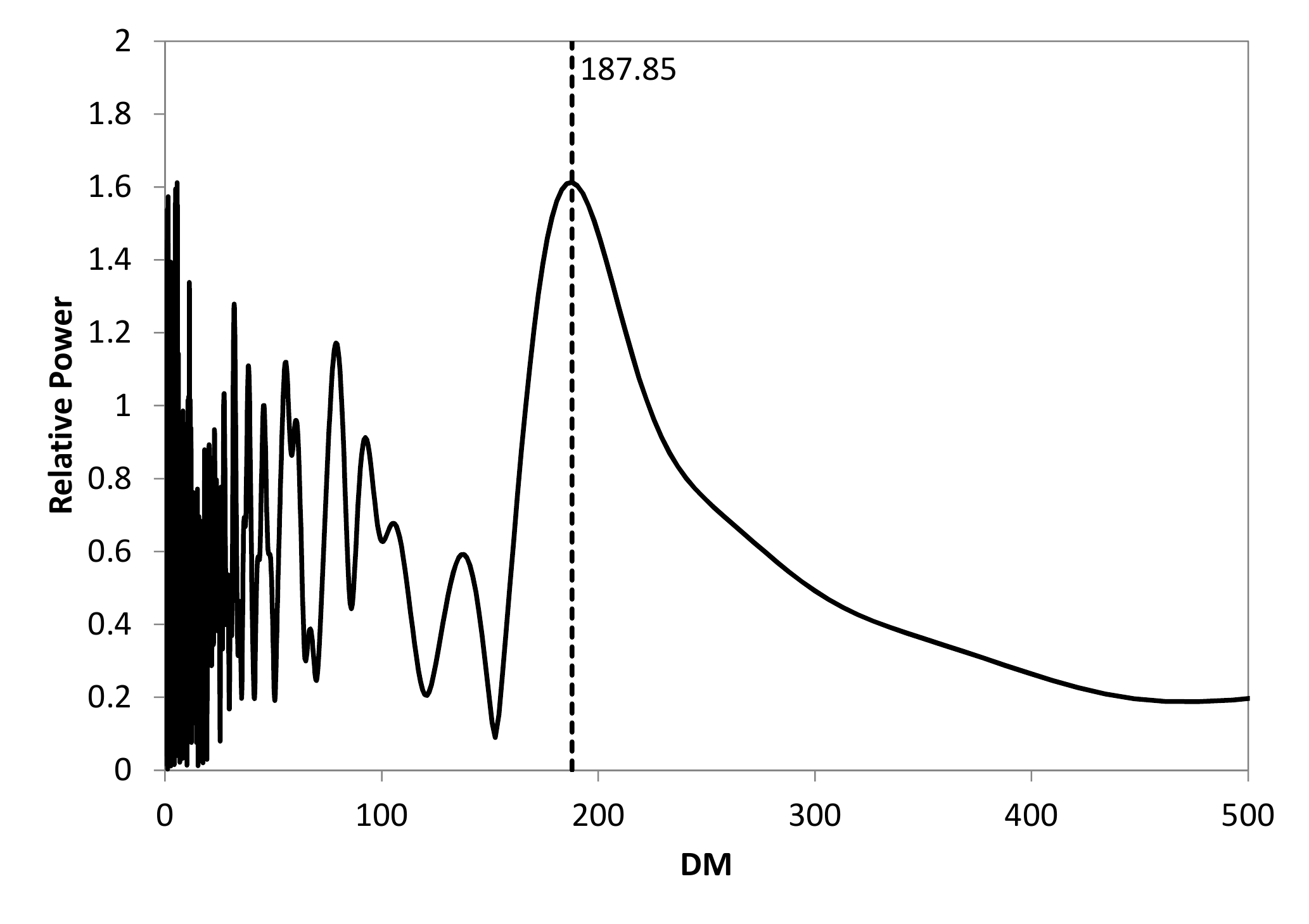}
\caption{\label{fig:periodo}Periodogram of DMs, showing highest peak at 187.85 cm$^{-3}$ pc.}
\end{figure}

Furthermore, one might try a periodogram as an independent peak search. As shown in Fig.~\ref{fig:periodo}, the highest peak is at 187.85 cm$^{-3}$ pc -- this is similar to the least-squares fit (187.7cm$^{-3}$ pc). 

We conclude that the trend is not likely to be the result of numerical fluctuations, but we remain cautious due to the problem of forming an unmotivated hypothesis after looking at the data. Examples of finding false clustering in sky map data in a slightly different context, such as cosmic ray arrival directions, are abundant; but these do get sorted out by further data.

\begin{figure}
\includegraphics[width=\linewidth]{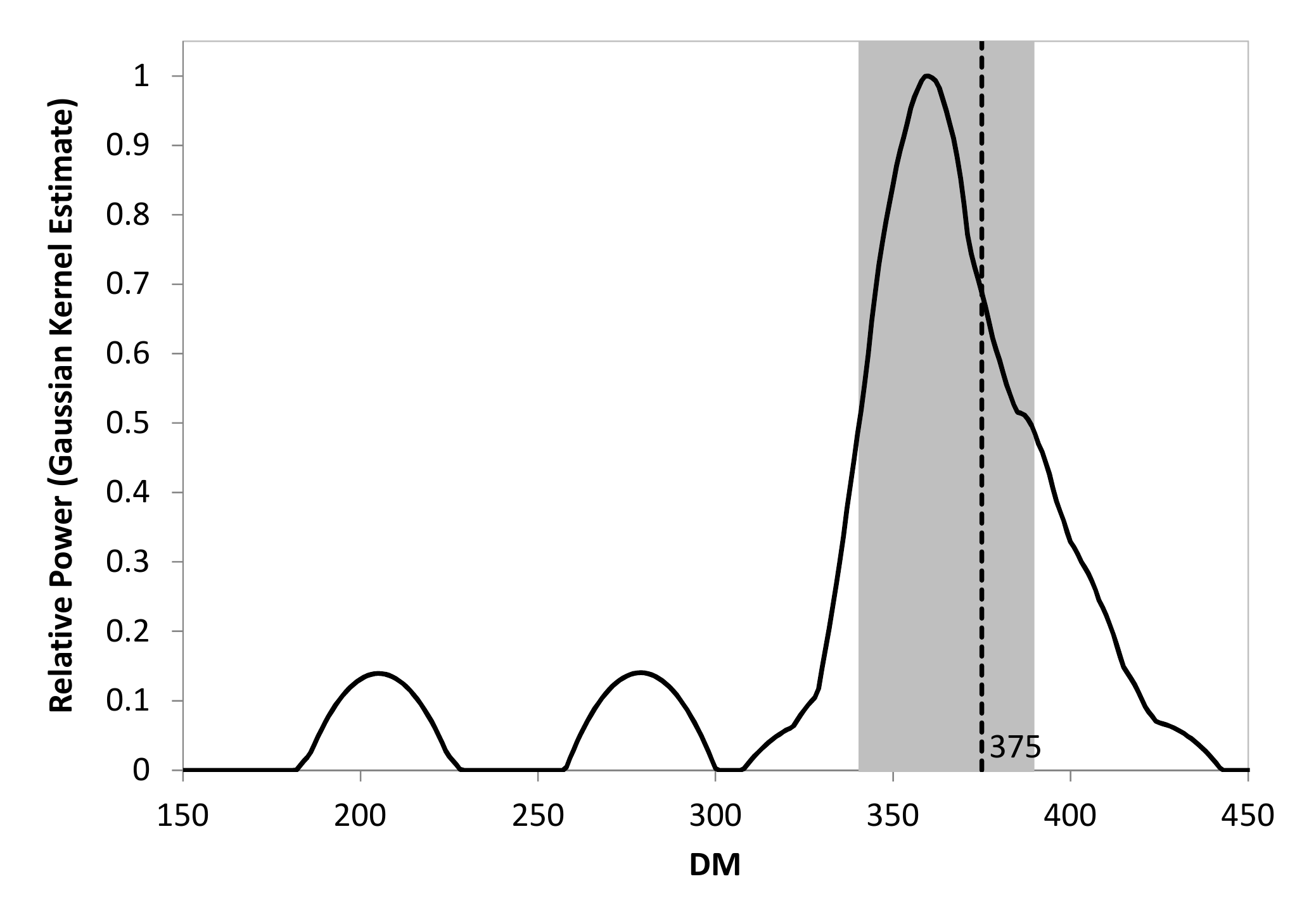}
\caption{\label{fig:perys}Peryton DM probability distribution shown as Gaussian Kernel Density Estimate. The full width at half mean of the highest peak (grey shade) covers the value for the lowest FRB found, 375 cm$^{-3}$ pc. This was already noted by \citet{Burke2011} and \citet{Kulkarni2014}.}
\end{figure}

\begin{figure}
\includegraphics[width=\linewidth]{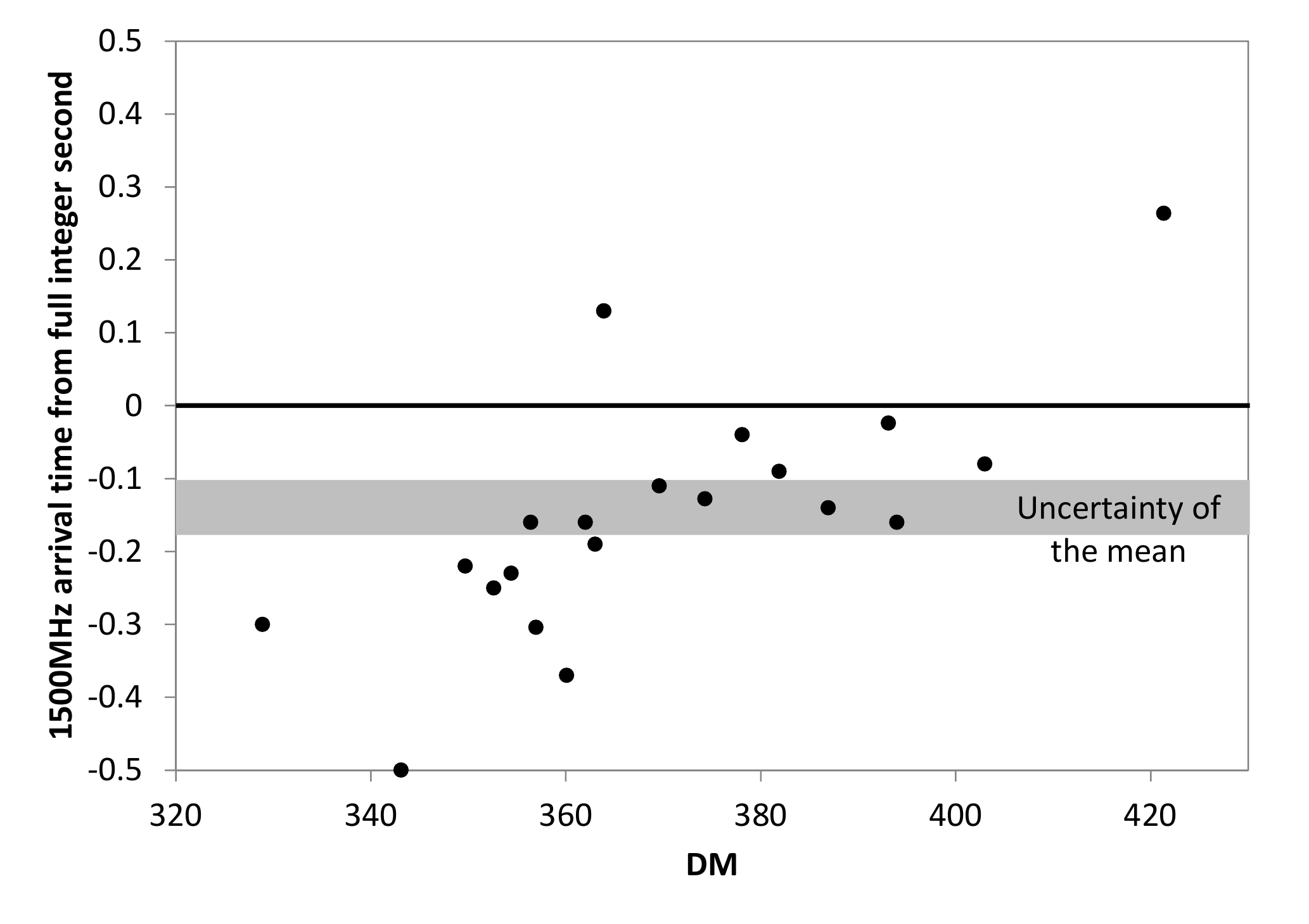}

\includegraphics[width=\linewidth]{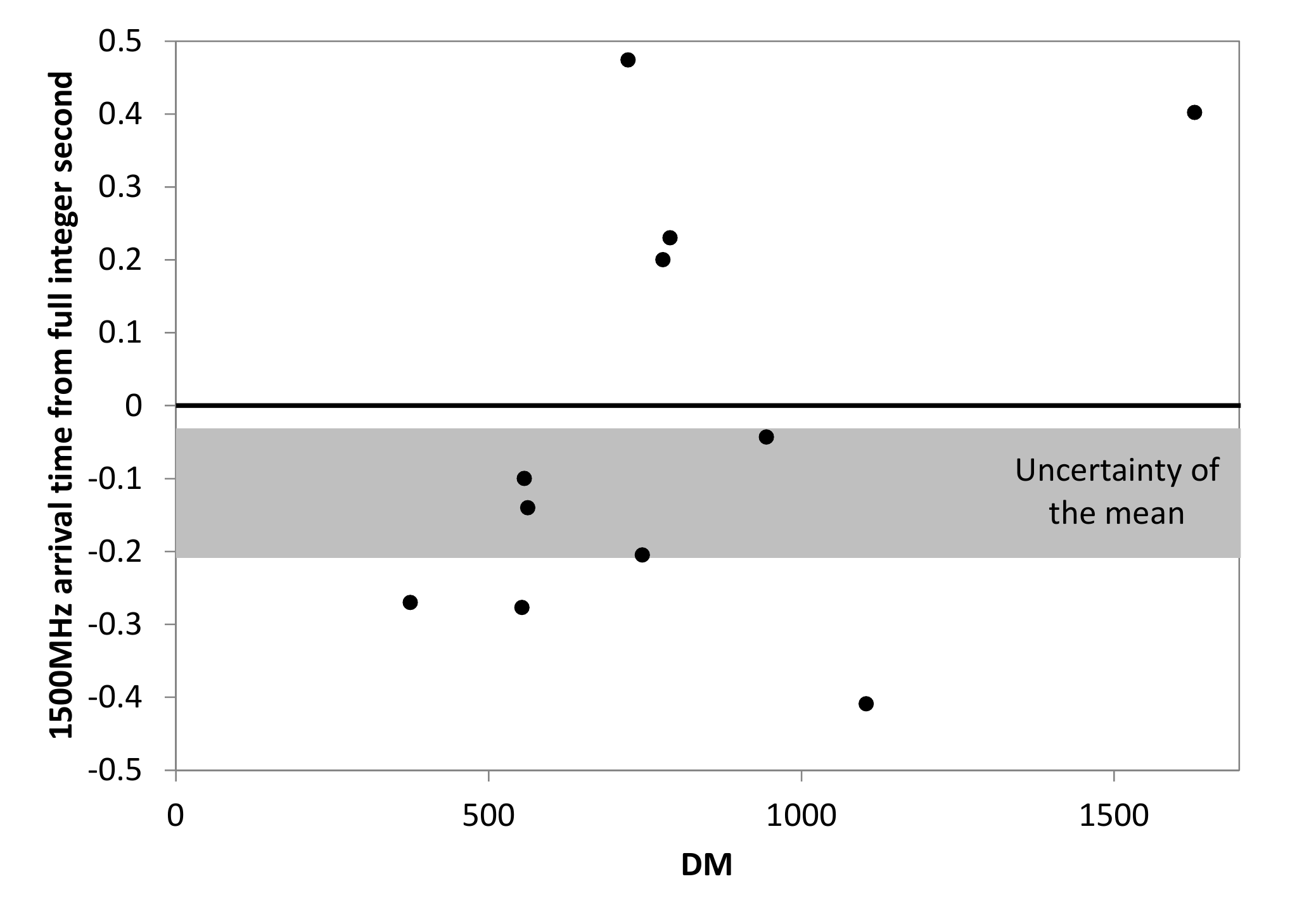}

\includegraphics[width=\linewidth]{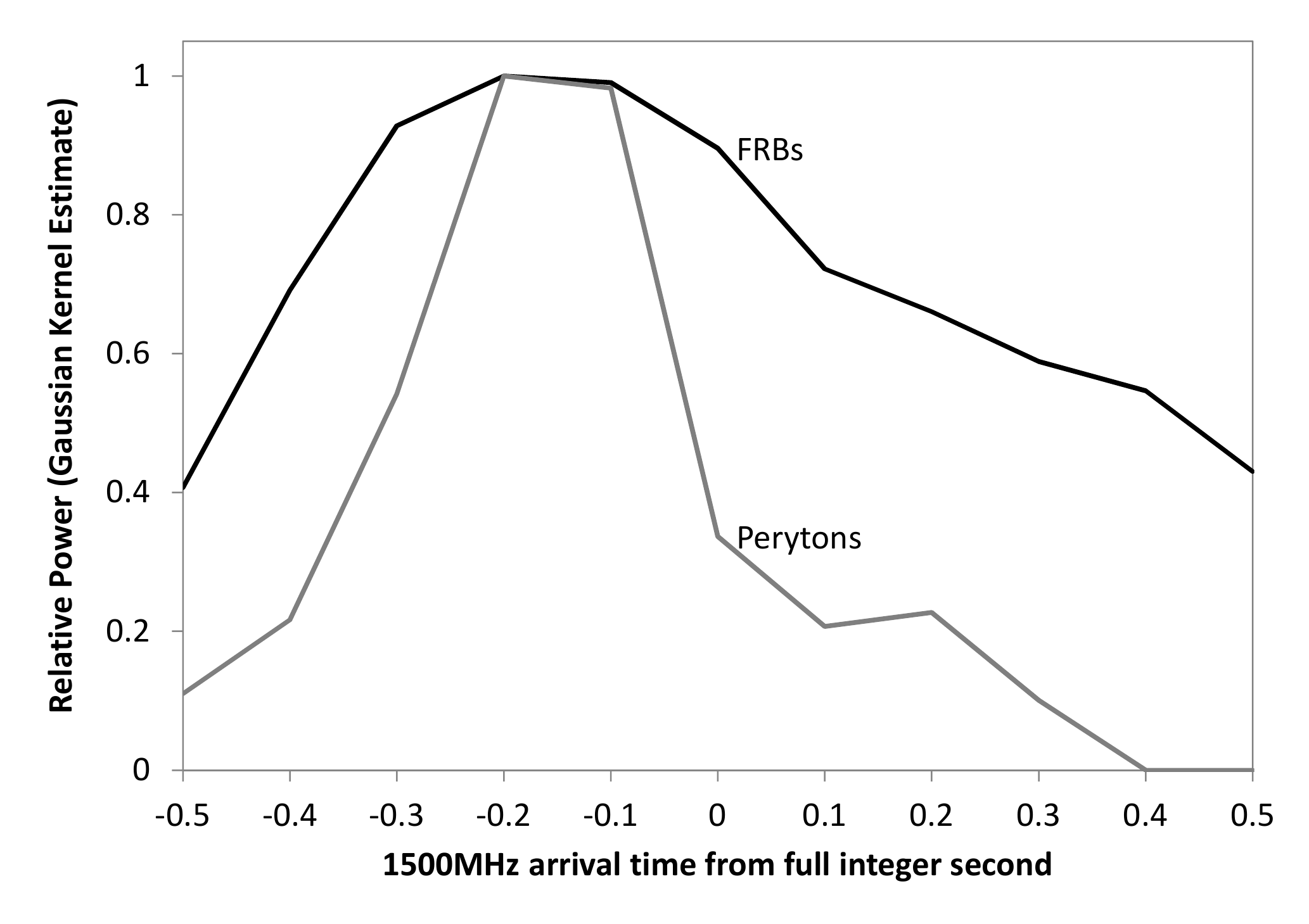}
\caption{\label{fig:time}Arrival time for the perytons (top, showing those with DM$\sim$375 cm$^{-3}$ pc and sub-second time resolution) and FRBs (middle) compared to the nearest full integer second. Bottom: Gaussian kernel estimate for perytons (grey) and FRBs (black) show common peak at $\sim$-0.1s before the full integer second at 1500MHz.}
\end{figure}

\subsection{Arrival times}
A different possible origin for FRBs would be a close similarity to Perytons, as discussed in \citet{Kulkarni2014}. Perytons are signals that are produced either in Earth's atmosphere, or by an artificial source. They are nearby, as they are seen in most beams of the telescopes. Their ocurrence at close to the full integer of the arrival second is a direct link to man-made interference. Yet, they share some similarity with FRBs, namely their dispersion measure. A total of 30 perytons have been found; at Parkes telescope (25, \citet{Burke2011,Bagci2012,Kocz2012}) and Bleien Observatory (5, \citet{Saint2014}). As can be seen in Fig.~\ref{fig:perys}, the perytons show a strong clustering around 360cm$^{-3}$, with the Kernel uncertainty encompassing the DM of the lowest FRB found, 375cm$^{-3}$. The mean for the 26 perytons in this peak is $368.5\pm8.2$cm$^{-3}$ (at $2\sigma$ confidence). It is interesting to note that the model deviations seem to be similar to the width of the peryton DM distribution. 

It has been noted by \citet{Burke2011} that perytons show an arrival time close to the full integer second (Fig.~\ref{fig:time}). We have repeated this analysis for the FRBs, with the caveat that we had to convert different timestamps (e.g. JD, BJD, UTC) and frequencies (we want one arrival time, but the bursts arrive with frequency-dependent time delay). We converted all measures to UTC and show the arrival times at 1500MHz in the middle panel of Fig.~\ref{fig:time}. The mean zeropoints of arrival times of perytons and FRBs are, within the errors, the same. The mean of arrival times of perytons is $-0.14\pm0.03$s to the integer second, FRBs arrive at $-0.12\pm0.08$s. Of course, this is only valid assuming that there is some cut-off for the frequency, e.g. at 1600MHz. Only then, a comparable arrival time due to different DMs can be calculated for the FRBs. We also caution that some error has been introduced by the conversions of frequencies and times; it will be helpful if the raw data of all FRBs were released, including their millisecond UTC arrival times for 1500MHz. Then, the analysis should be repeated for greater accuracy.

For both samples, we have performed a Kolmogorov-Smirnov-test, asking the question whether the distribution is uniform. This is rejected for perytons at the 0.1\% significance level, and for the FRBs at $p=0.03$, i.e. at the 95\% significance level.

\section{Discussion}
\label{sec:discussion}
The deviations from the model fit are summarized in Table~\ref{tab:dms}. We speculate that if the FRBs are indeed external to earth, then some of the deviations from DM linear spacing could be due to Milky Way contributions, which would be credible for $\sim$32--48cm$^{-3}$ pc in most cases \citep{Keane2015}. This would leave a large DM, in discrete steps, for the generating FRB process. We thus speculate that the producer of these bursts is located in our galaxy, and generates DMs in such discrete steps. We encourage modeling possible generator processes.

Following our theory of discrete spacings, one might argue that perytons are generated by the same process as FRBs, at an integer multiple of $2\times$187.5cm$^{-3}$. As perytons are thought to be produced on Earth, this would imply that FRBs are also Earthly noise. Indeed, why would both perytons and FRBs show arrival times with a strong correlation to Earth's integer second? This hints at some man-made device, such as mobile phone base stations. The device needs to keep (or sync) the time to sub-second precision, which is not common for e.g. microwave ovens.

\section{Conclusion}
\label{sec:conclusion}
We have noted a potential discrete spacing in DM of FRBs. Identified steps are integer multiples of 187.5cm$^{-3}$ pc, so that DMs occur in groups centered at 375, 562, 750, 937, 1125cm$^{-3}$ pc, with errors $<$5\%. If this holds, future FRBs would show DMs in these groups (and perhaps at the base 187.5cm$^{-3}$ pc, or larger integer multiples beyond 1125cm$^{-3}$ pc).

In case this would hold, an extragalactic origin would seem unlikely, as high (random) DMs would be added by intergalactic dust. A more likely option could be a galactic source producing quantized chirped signals, but this seems most surprising. If both of these options could be excluded, only an artificial source (human or non-human) must be considered, particularly since most bursts have been observed in only one location (Parkes radio telescope). A re-assessment of man-made phenomena, such as perytons \citep{Burke2011}, would then be required. Failing some observational bias, the suggestive correlation with terrestrial time standards seems to nearly clinch the case for human association of these peculiar phenomena.

In the end we only claim interesting features which further data will verify or refute.

\end{document}